\documentclass[twocolumn,prb,showpacs,preprintnumbers,amsmath,amssymb]{revtex4}

\usepackage{graphicx}
\usepackage{dcolumn}
\usepackage{bm}
\usepackage{epsfig}

\begin{document}


\title{Anharmonic soft modes in glasses}

\author{U. Buchenau}
\email{buchenau-juelich@t-online.de}
\affiliation{%
Forschungszentrum J\"ulich GmbH, J\"ulich Centre for Neutron Science (JCNS-1)
and Institute for Complex Systems (ICS-1), 52425 J\"ulich, Germany}%
\date{April 1, 2020}

\begin{abstract}
The anharmonic soft modes studied in recent numerical work in the glass phase of simple liquids have an unstable core, stabilized by the positive restoring forces of the surrounding elastic medium. The present paper formulates an unstable core version of the phenomenological soft potential model for the low temperature anomalies of glasses, relates a new numerical finding on low-barrier relaxations to old soft potential model results, and discusses experimental evidence for an unstable core of the boson peak modes. 
\end{abstract}


\maketitle

The low temperature properties of glasses around 1 K differ dramatically from those of simple crystals \cite{zepo}. Below 1 K, two level states, coupling strongly to external shear and compression, dominate the heat capacity and the thermal conductivity \cite{zepo,philbook}. Above 1 K, the vibrational spectrum shows a boson peak on top of the Debye density of sound waves \cite{zepo,bu86}.

There are successful empirical models like the tunneling model \cite{philtun} and its extension to include low barrier relaxations and soft vibrations, the soft potential model \cite{kki,bggprs,parshin,ramos,schober}. But what is tunneling or vibrating there, and why it couples so strongly to shear and compression, remains an open question.

The last four years brought two important new numerical developments. The first was a dedicated study of the localized vibrational soft modes in simple glasses \cite{le1,gale1,manning,le2,corein,mizuno,wang,le3} with the vibrational density of states $g(\omega)\propto\omega^4$ ($\omega$ frequency) predicted by the soft potential model \cite{bggprs,parshin,ramos,schober} and exhibiting the strong positive fourth order term $\chi_4$ in the mode potential \cite{le1,le2,le3} which the soft potential model needs for a common description of tunneling states and vibrations. The second was the swap mechanism for simple liquids \cite{swap}, which enables the numerical cooling of simple liquids down to temperatures which are even lower than the glass temperature of real liquids. The application of the swap mechanism to undercooled liquids revealed the central role of the soft localized modes for the understanding of the mode coupling transition \cite{coslovich} and documented a strong decrease of the number of these soft modes in the glass phase with decreasing glass temperature \cite{wang}.

One of these new papers \cite{corein} corroborated an important earlier numerical result \cite{corei}, namely the finding of an unstable core of the soft vibrational modes. The size of these unstable cores explodes as one approaches the jamming transition at the pressure zero \cite{corein}. The small positive force constant of the resulting soft vibration is due to the compensation of the negative restoring force of the core by the positive restoring force of the stable surroundings. 

The present paper formulates an unstable core version of the soft potential model, relates a new numerical finding on the barrier density of low-barrier relaxations \cite{le3} to old soft potential results \cite{ramos}, and explains the puzzling temperature dependence of the boson peak in the glass phase of silica \cite{wischi,wischidoc} and of two polymers \cite{schoenfeld} in terms of an unstable core of the boson peak modes. At the boson peak, the negative force constant of the core is overcompensated by the positive restoring Eshelby \cite{eshelby} forces from the surrounding elastic medium. The Eshelby forces are due to the elastic deformation of the unstable core as it moves along the eigenvector of the soft mode. 

The concept is taken from an Eshelby explanation of the boson peak in metallic glasses \cite{buscho}, in which two gliding triangles, a six-atom core, freeze in the middle between two close-packing configurations (an octahedron and two edge-sharing tetrahedra) by the elastic restoring forces from the surroundings. In this case, the core deformation is a pure shear (see Fig. 1).

\begin{figure}[b]
\hspace{-0cm} \vspace{0cm} \epsfig{file=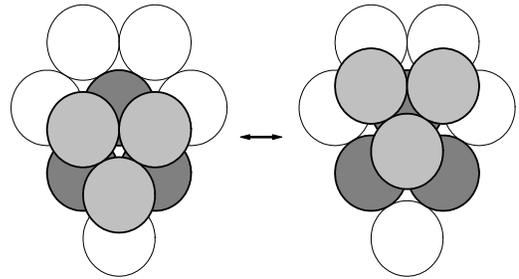,width=7cm,angle=0} \vspace{0cm}\caption{The gliding triangle shear transformation \cite{buscho} in simple liquids converts the six-atom octahedron on the left side into two edge-sharing tetrahedra on the right.} 
\end{figure}

Though the concept is plausible, the calculation \cite{buscho} contains a severe error, corrected recently \cite{buth}. It is correct that the creation energy $V_t$ of the gliding triangle soft mode is
\begin{equation}\label{vt}
	V_t=\frac{3GV_a}{2\pi^2},
\end{equation}
where $G$ is the shear modulus and $V_a$ is the atomic volume.

But the ratio $GV_a/k_BT_g$ at the glass transition with the glass temperature $T_g$ has been wrongly estimated \cite{buscho} to be 17.6 from the data collection \cite{johnson} for many metallic glasses. The real average ratio is 65.6, a factor of 3.73 larger.

Consequently, the creation energy of the gliding triangle soft mode is not 2.67 $k_BT_g$ as estimated \cite{buscho}, but rather 10 $k_BT_g$. Therefore it is no wonder that the real soft modes look differently.

The realization of the error led to the new proposal \cite{buth} for the core of the soft mode of Fig. 2, centered in the single-atom jump between two close-packed equilibrium positions. The unambiguous definition of the two close-packing positions requires seven atoms, a combination of an octahedron with a tetrahedron.

The picture explains the string-like motion in the core of the soft mode, observed first in numerical work on soft modes and low-barrier relaxations in the glass phase \cite{olig} and later in the flow processes at the glass transition \cite{kob} of simple liquids. The string runs along the line connecting the two atom positions, consisting of the single atom and its two nearest neighbors in the line direction, which the single atom pushes and pulls, respectively. With three atoms moving against seven atoms, it is clear that the string motion is much better visible.

The creation energy $V_s$ of such a string mode is one third of the one for a gliding triangle
\begin{equation}\label{vs}
	V_{s}=\frac{GV_a}{2\pi^2},
\end{equation}

\begin{figure}[b]
\hspace{-0cm} \vspace{0cm} \epsfig{file=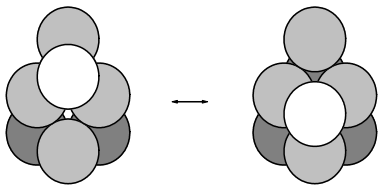,width=6 cm,angle=0} \vspace{0cm} \caption{A single-particle jump on the surface of a close-packed seven-particle unit (an octahedron joined to a tetrahedron) \cite{buth}.}
\end{figure}

With this lower barrier, it turned out to be possible to explain the reversible Kohlrausch tail of the viscous flow, with a Kohlrausch exponent of about one half, quantitatively \cite{buth}. 

The unstable core analysis \cite{buscho} of the gliding triangle shear transformation to derive the parameters of the tunneling and the soft potential model will be generalized here to cases where the configurational coordinate of the transformation of the core of the mode is no longer a pure strain. Following \cite{buscho}, we describe the two structural minima of the core by a cosine potential and the outer restoring forces by a quadratic term in the mode coordinate $y$
\begin{equation}\label{cos}
	V(y)=v_4\left(\cos{y}-1+\frac{y^2}{2}\right),
\end{equation}
where it is already assumed that the zero point of the outside forces coincides with the top of the core barrier and that one has perfect cancellation of the harmonic inside and outside terms, leading to the approximation
\begin{equation}
	V(y)=\frac{v_4}{24}y^4
\end{equation}
in the neighborhood of $y = 0$. Note that the creation energy of the mode is $2v_4$, the height of the frozen saddle point for the core potential.

The crossover energy $W$ between tunneling states and vibrations in the soft potential model is the zero point energy in this quartic potential, determined by the equality of potential energy and kinetic confinement energy. To calculate the kinetic confinement energy, one needs to know how many atoms move in the mode. To characterize the atomic motion, let us consider the potential minima of the core at $y_1,y_2 =\pm\pi$, and denote by $u_i$ the displacement of atom i with the atomic mass $M_i$ from one of these minima to the other. For a given $y$, the atomic displacement $\left|\vec{x}_i\right|=yu_i/2\pi$, so the normal coordinate $A$ of the mode (defined by the kinetic energy $\dot{A}^2/2$) is given by
\begin{equation}
	A^2=\frac{y^2}{4\pi^2}\sum_{i=1}^NM_iu_i^2,
\end{equation}
where the sum extends over all atoms in the sample. The normal coordinate relates the atomic displacement vector $\vec{x}_i$ to the normalized eigenvector $\vec{e}_i$ of the mode
\begin{equation}
	\vec{x}_i=\frac{A}{M_i^{1/2}}\vec{e}_i.
\end{equation}

According to the definitions of the soft potential model \cite{ramos}
\begin{equation}
	W\equiv\frac{4\pi^2}{2y_0^2\sum_{i=1}^NM_iu_i^2}=\frac{v_4}{24}y_0^4,
\end{equation}
where the left part defines $y_0$ by the equality of potential energy and kinetic confinement energy and leads to
\begin{equation}
	y_0=2^{3/4}3^{1/4}\frac{W^{1/4}}{v_4^{1/4}}
\end{equation}
and
\begin{equation}\label{w}
	W=\frac{\pi^{4/3}}{6^{1/3}}\left(\frac{\hbar^2}{\sum_{i=1}^NM_iu_i^2}\right)^{2/3}v_4^{1/3}.
\end{equation}

Of course, $W$ could be directly evaluated from the numerical data, without any approximation, because one only needs the eigenvector and the fourth order potential term for the eigenvector. The determination of the creation energy is not quite so easy, because one needs to determine the two energy minima of the frozen core. But it should still be attempted, because it is a central information. 

There is a very recent important numerical result \cite{le3}, showing that one has double well potentials with a barrier density proportional to $V_b^{1/4}$, where $V_b$ is the barrier height. The work is based on an earlier ingenious characterization \cite{gale1,gale2} of the soft modes in terms of eigenvectors defined over the fourth and third terms of the mode potential. This characterization led to the surprising result that the eigenvectors defined over the fourth order term are very close to the usual second order ones, allowing one to get rid of the influence of the hybridization between localized modes and phonons.

In the soft potential model, the double well potentials are due to modes with a negative restoring force $D_2$, as long as the absolute value of the linear potential coefficient $D_1$ stays below $\left|D_2\right|^{3/2}/3\sqrt{6}$. For a constant density of modes in the $D_1,D_2$-plane, this leads again to a barrier density increasing with $V_b^{1/4}$.

It has not yet been possible to check the validity of the prediction numerically, because it is not easy to search for neighboring energy minima in numerical work \cite{le3}. But the soft potential prediction has been checked many years ago \cite{ramos}, because it leads to a sound wave absorption
proportional to $T^{3/4}$ as soon as the classical relaxation starts to prevail over the absorption by tunneling states. This happens at the crossover temperature
\begin{equation}
	T_c=1.22\frac{W}{k_B},
\end{equation}
where the tunneling plateau ends and the $T^{3/4}$-rise begins. The $T^{3/4}$-rise is indeed found in vitreous silica, vitreous germania and in B$_2$O$_3$-glass \cite{ramos}. 

But it does not go very far out in temperature. There is a low temperature peak in the sound absorption, showing that the barrier distribution of the soft modes has a cutoff at a barrier height which is smaller than the thermal energy $k_BT_g$ at the glass transition. In polymethylmethacrylate already the tunneling plateau has a downward inclination, indicating a cutoff barrier close to zero \cite{nittke}.

In the opposite direction, toward the boson peak, the numerical data \cite{mizuno} show that the boson peak marks the point where the sound waves begin to disappear, leaving only extended modes which are a mixture of sound waves and local modes at higher frequencies. The question addressed here is whether the local parts of the boson peak modes are again unstable cores. This would imply that the boson peak modes do also have an inner unstable core, in this case with a negative inner force constant overcompensated by the external elastic Eshelby forces.

The existence of an unstable core in the boson peak modes can be demonstrated convincingly from the temperature dependence of the boson peak in some glasses. The best example is vitreous silica, where the core motion is a coupled rotation of many corner-connected SiO$_4$-tetrahedra, and the positive contribution from the outside is due to the elastic constants \cite{fabiani}. 

Fig. 3 shows the temperature dependence of the boson peak frequency $\omega_b$ in silica \cite{wischi,wischidoc} together with a fit which assumes that only the elastic constants depend on temperature, while the core retains a temperature-independent negative restoring force constant. The temperature dependence of the elastic constants is taken from Brillouin measurements \cite{vacher,dardy}. Note that the fit reproduces not only the strong rise of the boson peak frequency at elevated temperatures, but also the low-temperature minimum.

\begin{figure}[b]
\hspace{-0cm} \vspace{0cm} \epsfig{file=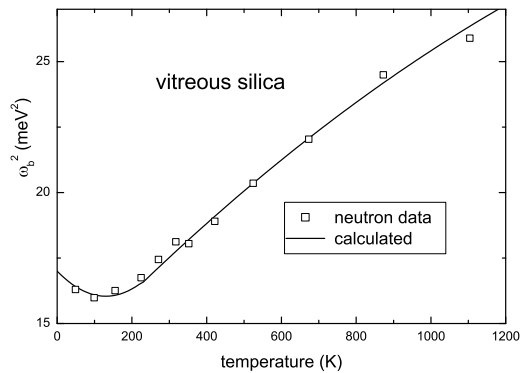,width=7cm,angle=0} \vspace{0cm}\caption{Temperature dependence of the boson peak frequency in silica glass \cite{wischi,wischidoc}, explained in terms of an unstable core of many SiO$_4$-tetrahedra, stabilized by the surrounding elastic medium.} 
\end{figure}

For the fit in Fig. 3, one needs to modify eq. (\ref{cos}) to the case where the outer force constant $v_{ext}$ overcompensates the negative inner one
\begin{equation}
	V(y)=v_4(\cos{y}-1)+v_{ext}\frac{y^2}{2}
\end{equation}
leading to a harmonic vibration with the boson peak frequency
\begin{equation}
	\omega_b^2=(v_{ext}-v_4)\frac{4\pi^2}{\sum_{i=1}^NM_iu_i^2}
\end{equation}
so one gets the relation
\begin{equation}
	\frac{\omega_b^2(T)}{\omega_b^2(0)}=\frac{v_{ext}(T)-v_4}{v_{ext}(0)-v_4}=\frac{G(T)-G_4}{G(0)-G_4}
\end{equation}
where $G(T)$ is the temperature-dependent shear modulus, assumed to be proportional to $v_{ext}(T)$. $G_4$ is proportional to $v_4$ and assumed to be independent of temperature, because it is a property of a local saddle point in a glass which does not change its structure with temperature. The fit in Fig. 3 requires $G_4=25$ GPa, while the shear modulus varies between 30.7 GPa at 145 K and 35 GPa at 1473 K. It demonstrates clearly that the boson peak modes in silica have an unstable core, with a negative restoring force constant which is seventy to eighty percent of the positive restoring forces from the elastic surroundings.

This interpretation collides with a recent one \cite{chuma}, which postulates that the density is the essential quantity determining the boson peak frequency, independent of whether one deals with a crystal or a glass. The conclusion was drawn from a comparison of the boson peaks in normal and densified silica with the lowest van-Hove singularities in quartz and cristobalite. But this conclusion is contradicted by the pronounced hardening of the boson peak frequency with increasing temperature in vitreous silica, which has a nearly temperature-independent density \cite{brueckner}. Also, it is a questionable conclusion in crystals where the lowest van-Hove singularity is connected with the soft mode of a structural phase transformation \cite{grimm}.

The opposite behavior to vitreous silica was observed for the boson peak in two polycarbonates \cite{schoenfeld}. The boson peak frequency decreases by a factor of two in the glass phase between 50 and 300 K, well below the glass temperature of about 420 K. But in this case the behavior of the shear modulus is also the opposite one: $G(T)$ decreases with increasing temperature \cite{patt,scie,fuku}, though only by twenty to thirty percent between 50 and 300 K, again much more weakly than the square of the boson peak frequency, which decreases by a factor of four. This indicates a temperature-independent negative core restoring force, with an absolute value of more than 90 percent of the positive external elastic restoring force.

To conclude, the soft potential model for the low temperature glass anomalies has been formulated in terms of an unstable core of the soft modes, a concept suggested by numerical results. The recently published numerical prediction of a barrier density $p(V_b)\propto V_b^{1/4}$ is once more in accordance with the soft potential model, corroborated by low temperature sound absorption data in real glasses. The numerical finding of an unstable core of the soft vibrational modes seems to extend up to the boson peak, because it supplies a convincing explanation for the puzzling temperature dependence of the boson peak in the glass phases of vitreous silica and two polymers.

Thanks are due to Edan Lerner and Herbert Schober for helpful suggestions, and to Miguel Angel Ramos for insisting on a return to the topic.

\end{document}